\documentclass[pra,aps,twocolumn,superscriptaddress,showpacs]{revtex4}
\usepackage{graphics}
\usepackage{bm}
\usepackage{amsmath}
\usepackage{amsfonts}
\usepackage{amssymb}
\usepackage{dsfont}

\begin{document}

\title{Loophole-free test of quantum non-locality using high-efficiency 
homodyne detectors}

\author{R. Garc\'{\i}a-Patr\'on\,}
\affiliation{QUIC, Ecole Polytechnique, CP 165, 
Universit\'{e} Libre de Bruxelles, 1050 Brussels, Belgium }

\author{J. Fiur\'{a}\v{s}ek}
\affiliation{QUIC, Ecole Polytechnique, CP 165, 
Universit\'{e} Libre de Bruxelles, 1050 Brussels, Belgium }
\affiliation{Department of Optics, Palack\'{y} University, 
17. listopadu 50, 77200 Olomouc, Czech Republic}

\author{N.J. Cerf\,}
\affiliation{QUIC, Ecole Polytechnique, CP 165, 
Universit\'{e} Libre de Bruxelles, 1050 Brussels, Belgium }

\begin{abstract}
We provide a detailed analysis of the recently proposed setup for a
loophole-free test of Bell inequality using conditionally generated
non-Gaussian states of light and balanced homodyning. In the proposed scheme, 
a two-mode squeezed vacuum state is de-gaussified by subtracting a single photon from each mode
with the use of an unbalanced beam splitter and a standard low-efficiency 
single-photon detector. We thoroughly discuss the dependence of the
achievable Bell violation on the various relevant experimental parameters 
such as the detector efficiencies, the electronic noise and the  mixedness 
of the initial Gaussian state.  We also consider several alternative schemes 
involving squeezed states, linear optical elements, conditional photon 
subtraction and homodyne detection. 

\end{abstract}

\maketitle

\section{Introduction}
In their seminal 1935 paper, Einstein, Podolsky, and Rosen (EPR) advocated 
that if ``local realism'' is taken for granted, then quantum theory is an incomplete 
description of the physical world \cite{Einstein35}. 
The EPR argument gained a renewed attention in 1964, 
when John Bell derived his famous inequalities, which must be satisfied
within the framework of any local realistic theory \cite{Bell64}. 
The violation of Bell inequalities, predicted by quantum mechanics, 
has since then been observed in many experiments 
\cite{Freedman72,Aspect81,Aspect82a,Aspect82b,Kwiat95,Weihs98,Rowe01,Tittel98},
thereby disproving the concept of local realism.
So far, however, all these tests suffered from 
either a detector-efficiency loophole or a locality loophole
\cite{Pearle70,Kwiat94}, that is,
the measured correlations may be explained in terms of local 
realistic theories exploiting the low detector efficiency
or the timelike interval between the two detection events 
\cite{Santos92,Gisin99,Massar02}.

A test of Bell inequality violation typically involves two distant parties,
Alice and Bob, who simultaneously carry out measurements on parts of 
a shared quantum system that is prepared in an entangled state. 
Both parties randomly and independently decide between one of two possible 
quantum  measurements $a_1, a_2$ and  $b_1,b_2$. 
To avoid the locality loophole, 
the measurement events (including the choice of the measurement) at 
Alice's and Bob's sites must be spacelike separated.
This suggests that optical systems are particularly suitable candidates for the test of Bell inequality
violations. The technology of generation of entangled states of photons is
very well mastered today \cite{Kwiat95} and the prepared entangled states can be distributed over
long distances via low-loss optical fibers \cite{Weihs98}. 
However, the currently available single-photon detectors suffer from a 
too low efficiency $\eta$, which
opens the so-called detector-efficiency loophole. This loophole has been closed in
a recent experiment with two trapped ions \cite{Rowe01}. 
However, the ions were held in a single trap, only several 
micrometers apart, so that the measurement events were not space-like separated.
It was suggested that two distant trapped ions can be
entangled via entanglement swapping by first preparing an entangled state 
of an ion and a photon on each side and then projecting the two photons on 
a maximally entangled singlet state \cite{Simon03,Feng03,Duan03,Browne03}. This 
technique could be used to close the locality loophole in 
the Bell test with trapped ions \cite{Simon03}. 
Very recently, the first step toward this goal, namely the entanglement between 
a trapped ion and  a photon emitted by the ion, has been observed 
experimentally \cite{Blinov04}. However, the entanglement swapping would 
require interference of two photons emitted by two different ions, 
which is experimentally very challenging.

An interesting alternative to the atom-based approaches
\cite{Simon03,Fry95,Freyberger96} is represented 
by all-optical schemes involving balanced homodyne detection, which can exhibit 
very high detection efficiency \cite{Polzik92,Grosshans03}. 
Unfortunately, the entangled two-mode squeezed state that can easily be
generated experimentally \cite{Ou92,Schori02,Bowen04}
cannot be directly employed to test Bell inequalities with homodyning.  
This state is described by a positive definite Gaussian Wigner function, 
which thus provides a local hidden variable  model that can explain all 
correlations established via quadrature measurements carried by balanced 
homodyne detectors. Similarly as in the case of purification of continuous variable
entanglement \cite{Eisert02,Fiurasek02,Giedke02,Browne03b,Eisert04}, one has to go beyond the class of Gaussian states and Gaussian
operations.  For instance, it is  possible to obtain a violation 
of Bell inequality  with Gaussian two-mode squeezed vacuum state by performing 
photon-counting measurements \cite{Banaszek98} or the rather 
abstract measurements  described in Refs. \cite{Chen02,Mista02,Filip02}. 
However, in contrast to balanced homodyning, these measurements are either 
experimentally infeasible or suffer from a very low detection efficiency.

In order to close the detection loophole by using homodyne detectors, 
it is necessary to employ highly non-classical non-Gaussian entangled 
state whose Wigner function is not positive definite.  
Several recent theoretical works indeed 
demonstrated that violation of Bell inequalities can be observed using balanced 
homodyning \cite{Gilchrist98,Gilchrist99,Munro99,Wenger03}, if
specific entangled light states such as pair-coherent states, squeezed
Schr\"{o}dinger cat-like states, or  specifically tailored finite superpositions of
Fock states, are available. However, no feasible experimental scheme is known that could
generate the states required in Refs. \cite{Gilchrist98,Gilchrist99,Munro99,Wenger03}.

Recently, we have shown that a very simple non-Gaussian state obtained from two-mode  
squeezed vacuum by subtracting a single photon from each mode  
\cite{Opatrny00,Cochrane02,Olivares03} can exhibit Bell violation 
with homodyning \cite{Sanchez04}.   
An essential feature of our proposal is that the photon subtraction can be
successfully performed with low-efficiency single-photon detectors, which renders 
the setup experimentally feasible. In fact, the basic building block 
of the scheme, namely the de-gaussification of a single-mode squeezed vacuum 
via single-photon subtraction, has been recently successfully 
implemented experimentally \cite{Wenger04}. 

In the present paper, we provide a thorough analysis of the scheme 
proposed in Ref. \cite{Sanchez04}. We present the details of 
the calculation of the Bell factor for a realistic setup that takes 
into account mixed input states, losses, added noise and imperfect 
detectors. Moreover, we shall also discuss several alternative schemes 
that involve the subtraction of one, two, three, or four photons. 
The present paper is organized as follows. In Section II, we describe 
the proposed experimental setup and we introduce the Bell-CHSH inequalities. 
We then provide a simple pure-state analysis of the scheme assuming
ideal detectors, which gives an upper bound on the achievable Bell violation.
In Section III, we present the mathematical description of a realistic setup 
with imperfect detectors, losses and noise. Besides the scheme where 
a single photon is subtracted on each side, we will also analyze 
a scheme where two photons are subtracted on each side. This latter 
scheme yields slightly higher Bell violation but only at the expense of a very 
low probability of state preparation. Several other schemes 
composed of squeezed state sources, linear optics, and photon subtraction, 
are discussed in Section IV. Finally, the conclusions are drawn in Section V.

\section{Feasible Bell test with homodyne detection}

\subsection{Proposed optical setup}

The conceptual scheme of the proposed experimental setup  is depicted in Fig. 1. 
A source  generates a two-mode squeezed vacuum state in modes A and B. 
This can be accomplished, e.g., by 
means of non-degenerate parametric amplification in a $\chi^{(2)}$ 
nonlinear medium or by generating two single-mode squeezed vacuum states 
and combining them on a balanced beam splitter. Subsequently, the state is de-gaussified
by conditionally subtracting a single photon from each beam.  A tiny part of each 
beam is reflected from a beam splitter BS$_A$ (BS$_B$) with a high transmittance T. 
The reflected portions of the beams impinge on single-photon detectors such as avalanche
photodiodes. A successful photon number subtraction is heralded by a click of each
photodetector PD$_A$ and PD$_B$ \cite{Olivares03}. In practice, the photodetectors 
exhibit a single-photon sensitivity but not a single photon resolution,  that is, 
they can distinguish the absence and presence of photons but cannot measure 
the number of photons in the mode. Nevertheless, this is not a problem here because in the
limit of high $T$, the most probable event leading to the click of a photodetector is
precisely that a single photon has been reflected from the squeezed beam on the beam
splitter. The probability of an event where two or more photons are subtracted from 
a single mode is smaller by a factor of $\approx 1-T$ and becomes totally negligible 
in the limit of $T\rightarrow 1$. Another important feature of the scheme is that 
the detector efficiency $\eta$ can be quite low because small $\eta$ only reduces 
the success rate of the conditional single-photon subtraction but it does not
significantly decrease the fidelity of this operation. These issues will be 
discussed in detail in Section III.

\begin{figure}[t]
\centerline{\resizebox{0.99\hsize}{!}{\includegraphics*{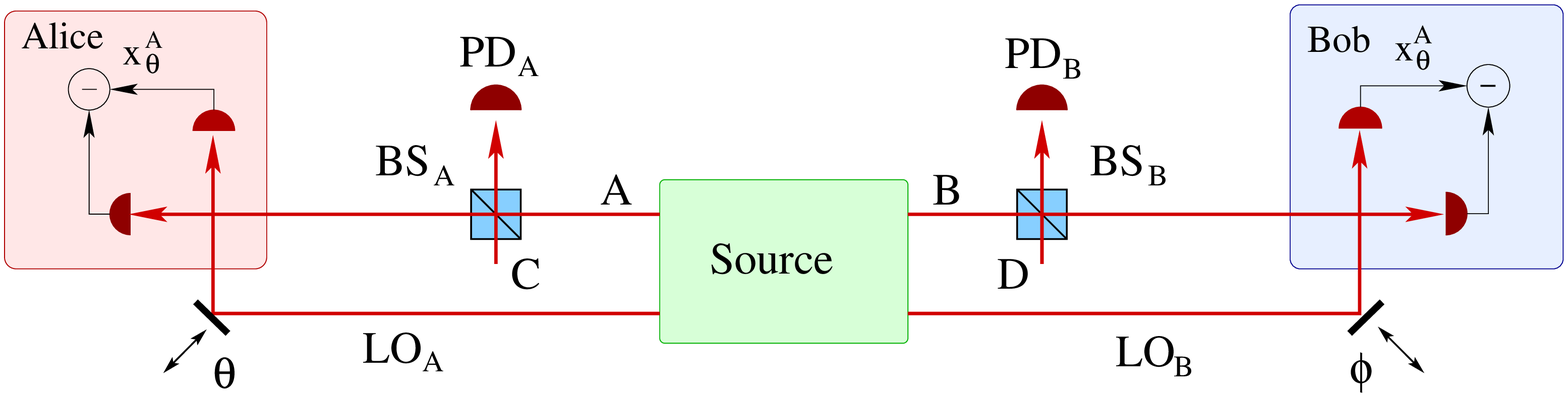}}}
\caption{Conceptual scheme of the proposed experimental setup for observing violation of
Bell inequalities with balanced homodyning. The source emits two-mode squeezed vacuum
mode in modes A and B. A small part of the beams is subtracted on two unbalanced beam
splitters BS$_A$ and BS$_B$ and sent on single-photon detectors 
PD$_A$ and PD$_B$.}
\end{figure}

After generation of the non-Gaussian state, the two beams A and B together with the
appropriate local oscillators LO$_A$ and LO$_B$ are sent to Alice and Bob, who
then randomly and independently measure one of two quadratures $x_{\theta_j}^A$, 
$x_{\phi_k}^B$ characterized by the relative phases $\theta_1,\theta_2$ and
$\phi_1,\phi_2$ between the measured beam and the corresponding local oscillator. 
The rotated quadratures $x_\theta^A=\cos\theta \, x^A+\sin\theta \, p^A$ and 
$x_\phi^B=\cos\phi \, x^B +\sin \phi \, p^B$ are defined in terms of the four quadrature
components of modes A and B  that satisfy the canonical commutation 
relations $[x^j,p^k]=i\delta_{jk}$, $j,k\in\{A,B\}$. 

To avoid the locality loophole, the whole experiment has to be carried out 
in the pulsed regime and a proper timing is necessary. In particular, the measurement
events on Alice's and Bob's sides (including the choice of phases) have to be space-like
separated. A specific feature of the proposed setup is that the non-Gaussian entangled 
state needed in the Bell test is generated conditionally when both ``event-ready'' 
detectors \cite{Bell} PD$_A$ and PD$_B$ click. However, we would like to stress
that this does not represent any loophole if proper timing is satisfied. Namely,
in each experimental run, the detection of the clicks (or no-clicks) of photodetectors 
PD$_A$ and PD$_B$ at the source should be be space-like separated from Alice's 
and Bob's measurements. This guarantees that the choice of the measurement basis
on Alice's and Bob's sides cannot in any way influence 
the conditioning measurement or vice versa  \cite{Sanchez04,Simon03,Bell}).

To demonstrate that the experimental data recorded by Alice and Bob are
incompatible with the concept of local realism, we shall consider the 
Bell-CHSH inequality originally devised for two-qubit system \cite{CHSH}. In this scenario, 
Alice (Bob) randomly and independently decides between one of two possible 
quantum measurements $a_1, a_2$ ($b_1, b_2$) which should have only  
two possible outcomes $+1$ or $-1$. We define the Bell parameter 
\begin{equation}
S=\langle a_1 b_1\rangle+\langle a_1b_2\rangle+\langle a_2b_1\rangle-\langle a_2b_2\rangle,
\label{S}
\end{equation}
where $\langle a_j b_k\rangle$ denotes the average over the subset of 
experimental data where Alice measured $a_j$ and, simultaneously, Bob measured $b_k$.
If the observed correlations can be explained within the framework of the 
local-hidden variable theories, then $S$ must satisfy the Bell-CHSH inequality $|S|\leq 2$.

In the proposed experiment, Alice and Bob measure quadratures which have continuous
spectrum. We discretize the quadratures  by postulating that the outcome is  $+1$
when $x\geq 0$ and $-1$ otherwise. The two different measurements on each side
correspond to the choices of two relative phases $\theta_1,\theta_2$ and $\phi_1,\phi_2$.
Quantum mechanically, the correlation $E(\theta_j,\phi_k)\equiv\langle a_j b_k\rangle$
can be expressed as  
\begin{equation}
E(\theta_j,\phi_k)=\int_{-\infty}^\infty \mathrm{sign}(x_{\theta_j}^A x_{\phi_k}^B)
P(x_{\theta_{j}}^A,x_{\phi_{k}}^B) d x_{\theta_{j}}^A d x_{\phi_{k}}^B,
\label{E}
\end{equation}
where $P(x_{\theta_j}^A,x_{\phi_k}^B)\equiv\langle
x_{\theta_j}^A,x_{\phi_k}^B|\rho_{c,AB}|x_{\theta_j}^A,x_{\phi_k}^B\rangle$ is the joint probability
distribution of the two commuting quadratures  $x_{\theta_{j}}^A$ and 
$x_{\phi_{k}}^B$, and $\rho_{c,AB}$ denotes the (normalized) conditionally generated 
non-Gaussian state of  modes A and B. In practice, the correlations would be determined
from the subset of the experimental data corresponding to the successful conditional
de-gaussification, i.e., Alice and Bob would discard all results obtained 
in measurement runs where either PD$_A$ or PD$_B$ did not click. We emphasize
again that this does not open any loophole in the Bell test.

\subsection{Ideal photodetectors}

We shall first present a simplified description of the setup, assuming 
ideal photodetectors ($\eta_{\mathrm{PD}}=1$) with single-photon resolution and  
conditioning on detecting exactly a single photon at each detector 
\cite{Opatrny00,Cochrane02}. This idealized treatment is valuable since it
provides an upper bound on the practically achievable Bell factor $S$. 
Moreover, as noted above, in the limit of high transmittance 
of $BS_A$ and $BS_B$, $T\rightarrow 1$, the 
realistic (inefficient) detector with single-photon sensitivity is in our case 
practically equivalent to these idealized detectors.

\begin{figure}[t]
\centerline{\resizebox{0.9\hsize}{!}{\includegraphics*{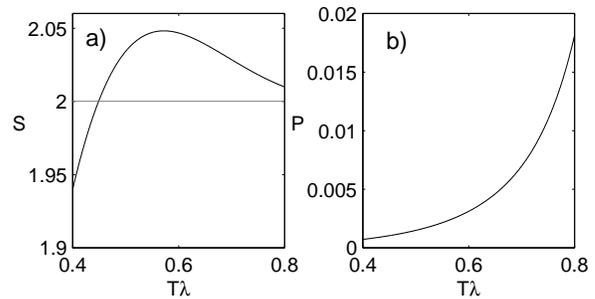}}}
\caption{  (a) Bell factor $S$ is plotted as a function of
 the effective squeezing parameter $T\lambda$ for $\theta_1=0$,
 $\theta_2=\pi/2$, $\phi_1=-\pi/4$ and $\phi_2=\pi/4$. (b) Probability $P$ 
 of successful conditional generation of the state
 $|\psi_{\mathrm{out}}\rangle$ as a function of the effective squeezing parameter
 $T\lambda$, assuming $T=0.95$.}
\end{figure}

The two-mode squeezed vacuum state can be expressed in the Fock state basis as 
follows,
\begin{equation}
|\psi_{\mathrm{in}}(\lambda)\rangle_{AB}=\sqrt{1-\lambda^2}\sum_{n=0}^\infty \lambda^n
|n,n\rangle_{AB},
\label{psiin}
\end{equation}
where $\lambda=\tanh(s)$ and $s$ is the squeezing constant. In the case of ideal
photodetectors, the photon number subtraction results in the state 
$|\psi_{\mathrm{out}}\rangle \propto a_{A}a_{B}|\psi_{\mathrm{in}}(T\lambda)\rangle$, where 
$a_{A,B}$ are annihilation operators and the parameter $\lambda$ is replaced by
$T\lambda$ in order to take into account the transmittance of $BS_A$ and $BS_B$.
A detailed calculation yields 
\begin{equation}
|\psi_{\mathrm{out}}\rangle_{AB}=\sqrt{\frac{(1-T^2\lambda^2)^3}{1+T^2\lambda^2}}
\sum_{n=0}^\infty (n+1)(T\lambda)^n|n,n\rangle_{AB},
\label{psiout}
\end{equation}
and the probability of the conditional preparation of state 
(\ref{psiout}) can be expressed as
\begin{equation}
\mathcal{P}=(1-T)^2\lambda^2(1-\lambda^2)\frac{1+T^2\lambda^2}{(1-T^2\lambda^2)^3}.
\label{Psimple}
\end{equation}

For pure states exhibiting perfect photon-number
correlations, the correlation coefficient (\ref{E}) depends only on the 
sum of the angles, $E(\theta_j,\phi_k)=\mathcal{E}(\theta_j+\phi_k)$.
With the help of the general formula derived by Munro \cite{Munro99} we obtain 
for the  state (\ref{psiout})
\begin{eqnarray}
\mathcal{E}(\varphi)&=& \frac{(1-T^2\lambda^2)^3}{1+T^2\lambda^2} \sum_{n>m}
\frac{8  \pi  (2T\lambda)^{n+m} }{n! m! (n-m)^2}(n+1)(m+1)
\nonumber \\
& &\times [\mathcal{F}(n,m)-\mathcal{F}(m,n)]^2 \cos[(n-m)\varphi],
\label{Esum}
\end{eqnarray}
where $\mathcal{F}(n,m)=\Gamma^{-1}((1-n)/2)\Gamma^{-1}(-m/2)$ and $\Gamma(x)$
stands for the Euler gamma function.

We have numerically
optimized the angles $\theta_{1,2}$  and $\phi_{1,2}$ to maximize the Bell 
factor $S$. It turns out that for any $\lambda$, it is optimal to choose $\theta_1=0$,
$\theta_2=\pi/2$, $\phi_1=-\pi/4$ and $\phi_{2}=\pi/4$. 
The Bell factor $S$ for this optimal choice of angles  is plotted as a function of 
the effective parameter $T\lambda$ in Fig. 2(a), and the corresponding probability 
of success of the conditional preparation of the state  
$|\psi_{\mathrm{out}}\rangle$ is plotted in Fig. 2(b).  We can see that $S$ is 
higher than $2$ so the Bell inequality is violated when $T\lambda>0.45$.
The maximal violation is achieved for $T\lambda\approx 0.57$, giving $S \approx 2.048$. 
This figure is quite close to the maximum Bell factor $S=2.076$ that could be reached
with homodyne detection, sign binning, and arbitrary states exhibiting perfect photon-number
correlations $|\psi\rangle=\sum_n c_n|n,n\rangle$ \cite{Munro99}.

\section{Realistic model}

In this section we will consider a realistic scheme with inefficient 
($\eta_{\mathrm{PD}}<1$) photodetectors exhibiting single photon sensitivity 
but no single-photon resolution, and realistic homodyning with 
efficiency $\eta_{\mathrm{\mathrm{BHD}}}<1$. The mathematical description of 
this  realistic model of the proposed experiment becomes strikingly simple
if we work in the phase-space representation and use the Wigner function formalism.
Even though the state used to test Bell inequalities is non-Gaussian, it can be 
expressed as a linear combination of four Gaussian states, so all the 
powerful Gaussian tools can still  be used. 

This section is further divided into three sub-sections. The first one
gives a brief overview of the Gaussian states, linear canonical transformations of
quadrature operators, and Gaussian completely positive maps.  In the second subsection,
an analytical formula for the Bell factor $S$ is derived, and the influence of 
detector inefficiencies, losses, and noise on the proposed Bell  experiment is
investigated in detail. Finally, an extended setup involving 
two-photon subtraction from each mode is studied in the third subsection.

\subsection{Gaussian states and Gaussian operations}

In quantum optics Gaussian states are often encountered as states of $n$ modes of light.
These states are completely specified by the first and second moments of the quadrature 
operators $r_k$ with $r=(x_1,p_1,.....,x_n,p_n)^T$. Here $r_k$ satisfy the 
canonical commutation relations (CCR) $[x_j,p_k]=i\delta_{j,k}$. 
Instead of referring to the density matrix one may refer to the 
Wigner function defined on phase space
\begin{equation}
W=\frac{1}{\pi^n\sqrt{\det\gamma}}\exp[-(r-d)^T\gamma^{-1} (r-d)],
\end{equation}
where $d$ is the vector of first moments, $d_j=\langle r_j \rangle$, 
and $\gamma$ is the covariance matrix
\begin{equation}
\gamma_{i,j}=\langle r_i r_j + r_j r_i\rangle-2d_i d_j. 
\end{equation}
In this paper we shall deal only with states with zero displacement $d_j=0$.
Some relevant examples of Gaussian states that we shall need in what follows include:
(i) The $n$-mode vacuum state with $d_j=0$ and covariance matrix equal to the identity
matrix, $\gamma_{\mathrm{Vac}}=I_{2n}$. (ii) Single-mode squeezed vacuum state 
with $d_j=0$ and covariance matrix 
\begin{equation}
\gamma_{\mathrm{SMS}}=\left [\begin{array}{cc}
e^{2s} & 0 \\
0 & e^{-2s} \\
\end{array}\right ],
\end{equation}
where $s$ is the squeezing parameter. 
(iii) Two-mode squeezed vacuum state with $d_j=0$ and covariance matrix
\begin{equation}
\gamma_{\mathrm{TMS}}=\left [\begin{array}{cccc}
\cosh(2s) & 0        & \sinh(2s) & 0         \\
0        & \cosh(2s) & 0        & -\sinh(2s) \\
\sinh(2s) & 0        & \cosh(2s) & 0         \\
0        & -\sinh(2s)& 0        & \cosh(2s)  \\
\end{array}\right ].
\label{gammaTMS}
\end{equation}

Optical operations that can be implemented with beam splitters, phase shifters, 
squeezers and homodyne detection correspond to Gaussian operations. Their important
property is that they map a Gaussian input state onto a Gaussian output state. 
Gaussian unitary transformations realize the mapping $r \rightarrow r'=Sr$
which preserves the CCR. This is the case if $S\in Sp(2n,\mathds{R})$ 
the so-called real symplectic group. On the covariance matrix level the 
transformation reads
\begin{equation}
\gamma \rightarrow S\gamma S^T.
\end{equation}
A particular subset of symplectic transformations is formed by the symplectic matrices
$S$ that are also orthogonal $S\in Sp(2n,\mathds{R})\cap O(2n)$. 
Those transformations are called passive  because they do not change the total number 
of photons.  The most common passive transformations include (i) 
mixing two modes of light with a beam splitter of (intensity) transmittance $T$ 
and reflectance $1-T$
\begin{equation}
S_{\mathrm{BS}}=\left [\begin{array}{cccc}
\sqrt{T}    & 0          & \sqrt{1-T} & 0           \\
0           & \sqrt{T}   & 0          & \sqrt{1-T}  \\
-\sqrt{1-T} & 0          & \sqrt{T}   & 0           \\
0           & -\sqrt{1-T} & 0         & \sqrt{T}  \\
\end{array}\right ],
\end{equation}
and a phase shift of a single mode
\begin{equation}
S_{\mathrm{PS}}(\theta)=\left [\begin{array}{cccc}
\cos\theta & \sin\theta            \\
-\sin \theta           & \cos\theta  \\
\end{array}\right ].
\end{equation}
All passive linear canonical transformations of $n$ modes can be implemented by optical
interferometers consisting of beam splitters and phase shifters. 

The second group of linear canonical transformations are the active transformations
that describe phase sensitive amplification of light. The archetypal examples are 
the single-mode squeezer
\begin{equation}
S_{\mathrm{SMS}}=\left [\begin{array}{cc}
e^{s} & 0 \\
0 & e^{-s} \\
\end{array}\right ]
\end{equation}
and the two-mode squeezer
\begin{equation}
S_{\mathrm{TMS}}=\left [\begin{array}{cccc}
\cosh(s)   & 0        & \sinh(s)  & 0        \\
0         & \cosh(s)  & 0        & -\sinh(s)  \\
\sinh(s)  & 0        & \cosh(s)  & 0        \\
0         & -\sinh(s) & 0        & \cosh(s)  \\
\end{array}\right ]. 
\end{equation}
These matrices describe the operation of ideal degenerate ($S_{\mathrm{SMS}}$) 
or nondegenerate ($S_{\mathrm{TMS}}$) optical parametric amplifier (OPA). 
In particular, a nondegenerate OPA provides a source of entanglement since it 
transforms the input vacuum into a two-mode squeezed vacuum state.

Noisy channels and phase-insensitive amplifiers are irreversible quantum operations 
which cannot be described  by Gaussian unitary transformations. Instead, they can be 
modeled within the more general framework of trace-preserving 
Gaussian completely positive (CP) maps \cite{Eisert02,Fiurasek02}. 
The covariance matrix transformation reads
\begin{equation}
\gamma \rightarrow A\gamma A^T+G.
\end{equation} 
Of particular importance is the propagation through a lossy quantum channel with
transmittance $\eta$, which is characterized by $A=\sqrt{\eta} \, I$ and $G=(1-\eta)I$.
In what follows, we shall use lossy channels followed by perfect detectors
to model inefficient detectors.

\subsection{Two photon subtractions}

We shall now present a detailed calculation of the Bell factor for the setup 
depicted in Fig. 1. Our model takes into account 
realistic photodetectors ($\eta_{\mathrm{PD}}<1$) with single-photon sensitivity, imperfect
homodyning and the added electronics noise.

\subsubsection{Preparation of a non-Gaussian state}

As shown in Fig. 1, the modes A and B 
are initially prepared in a two-mode squeezed vacuum state, 
and the auxiliary modes C and D are in vacuum state. The Wigner function of the
four-mode state ABCD is a Gaussian centered at the origin,
\begin{equation}
W_{\mathrm{in},ABCD}=\frac{1}{\pi^4\sqrt{\det\gamma_{\mathrm{in}}}}
\exp\left[-r^T \gamma^{-1}_{\mathrm{in}}r\right],
\label{Wignerin}
\end{equation}
where $r=[x^A,p^A,\ldots,x^D,p^D]$. The initial state is 
 fully characterized by the covariance matrix 
\begin{equation}
\gamma_{\mathrm{in}}=\gamma_{\mathrm{TMS},AB}\oplus I_{CD},
\end{equation}
where $\gamma_{\mathrm{TMS}}$ is the covariance matrix of a two-mode squeezed vacuum
(\ref{gammaTMS}) and $\oplus$ denotes the direct sum of matrices.

The imperfect single-photon detectors (balanced homodyne detectors) with detector
efficiency $\eta_{\mathrm{PD}}$ ($\eta_{\mathrm{BHD}}$) are modeled 
as a sequence of a lossy channel with transmittance $\eta_{\mathrm{PD}}$ 
($\eta_{\mathrm{BHD}}$) followed by an ideal photodetector (homodyne detector). 
In our setup, the  modes AC (BD)  interfere on the unbalanced 
beam splitters $BS_A$ ($BS_B$) and pass through the four ``virtual'' lossy channels 
before impinging on ideal detectors.  The  covariance matrix of 
the mixed Gaussian state  $\rho_{\mathrm{out},ABCD}$ just in front of the 
(ideal) detectors is related to $\gamma_{\mathrm{in}}$ via a Gaussian CP  map,
\begin{equation}
\gamma_{\mathrm{out}}=S_{\eta}S_{\mathrm{mix}}\gamma_{\mathrm{in}}S_{\mathrm{mix}}^T S_{\eta}^T+G,
\label{gammaout}
\end{equation}
where
\begin{eqnarray}
S_{\eta}=\sqrt{\eta_{\mathrm{BHD}}}I_{AB}\oplus \sqrt{\eta_{\mathrm{PD}}}I_{CD}, \\
G=(1-\eta_{\mathrm{BHD}})I_{AB}\oplus (1-\eta_{\mathrm{PD}})I_{CD},
\label{G}
\end{eqnarray}
and the symplectic matrix
\begin{equation}
S_{\mathrm{mix}}=S_{BS,AC} \oplus S_{BS,BD}
\end{equation}
describes the mixing of modes $A$ with $C$ and $B$ with $D$
on the unbalanced beam splitters BS$_A$ and BS$_B$, respectively. 

The state $\rho_{c,AB}$ is prepared by conditioning on observing clicks at both
photodetectors PD$_A$ and PD$_B$. These detectors respond with tho different outcomes, 
either a click, or no click. Mathematically, an ideal detector with a single 
photon sensitivity is described by a two-component positive operator valued
measure (POVM) consisting of the projectors onto the vacuum state and 
on the rest of the Hilbert space,
$\Pi_{0}=|0\rangle\langle 0|$, $\Pi_{1}=I-|0\rangle\langle 0 |$.
The resulting conditionally prepared state $\rho_{c,AB}$ can be calculated 
from the density matrix $\rho_{\mathrm{out}, ABCD}$ as follows,
\begin{equation}
\rho_{c,AB}=\mathrm{Tr}_{CD}[\rho_{\mathrm{out},ABCD} (I_{AB}\otimes
\Pi_{1,C}\otimes\Pi_{1,D})].
\label{rhocond}
\end{equation}
It is instructive to rewrite the partial trace in Eq. (\ref{rhocond}) in terms 
of Wigner functions, taking into account that
\begin{equation}
\mathrm{Tr}[XY]=(2\pi)^N\int_{-\infty}^\infty
W_X(r)W_Y(r) d^{2N} r, 
\end{equation}
where $W_X(r)$ and $W_Y(r)$ denote the Wigner representations of the 
operators $X$ and $Y$, respectively, and $N$ is the number of modes we trace over.
The POVM element $\Pi_1$ is a difference of two operators whose Wigner
representations are both Gaussian functions, $W_I=1/(2\pi)$,
$W_{0}=\pi^{-1}e^{-x^2-p^2}$. After a bit lengthy but otherwise straightforward
calculations we find that the Wigner function $W_{c,AB}$ 
of (normalized) conditionally prepared state (\ref{rhocond}) can be 
expressed as a linear combination of four Gaussian functions, 
\begin{equation}
W_{c,AB}(r)=\frac{1}{\pi^2 P_{G} \sqrt{\det\gamma_{\mathrm{out}}} }
\sum_{j=1}^4 \frac{q_j}{\sqrt{\det \Gamma_{j,CD}}}
e^{-r^T\Gamma_{j,AB}r},
\label{Wignercond}
\end{equation}
where $q_1=1$, $q_2=q_3=-2$ and $q_4=4$. 
The corresponding probability of success is given by
\begin{equation}
P_G=\frac{1}{\sqrt{\det\gamma_{\mathrm{out}}} }
\sum_{j=1}^4 \frac{q_j}{\sqrt{\det (\Gamma_{j,AB}\Gamma_{j,CD})}}.
\label{Proba}
\end{equation}
To define the various matrices appearing 
in Eqs. (\ref{Wignercond}) and (\ref{Proba}), we first introduce a matrix
$\Gamma=\gamma_{\mathrm{out}}^{-1}$ and we divide $\Gamma$ into four smaller submatrices
with respect to the bipartite $AB$ vs $CD$ splitting,
\begin{equation}
\Gamma=\left[
\begin{array}{cc}
\Gamma_{AB} & \sigma \\
\sigma^T & \Gamma_{CD}
\end{array}
\right].
\end{equation}
It holds that
\begin{equation}
\Gamma_{j,AB}=\Gamma_{AB}-\sigma \Gamma_{j,CD}^{-1}\sigma^T,
\label{intgauss}
\end{equation}
and the four matrices $\Gamma_{j,CD}$ read
\begin{equation}
\begin{array}{l}
\Gamma_{1,CD}=\Gamma_{CD},  \\
\Gamma_{2,CD}=\Gamma_{CD}+I_{C}\oplus 0_D,  \\
\Gamma_{3,CD}=\Gamma_{CD}+0_{C}\oplus I_D,  \\
\Gamma_{4,CD}=\Gamma_{CD}+I_{CD}.
\end{array}
\end{equation}

\subsubsection{Correlation coefficient $E(\theta_j,\phi_k)$}

The joint probability distribution $P(x_{\theta_{j}}^A,x_{\phi_{k}}^B)$ of the quadratures 
$x_{\theta_j}^A$ and $x_{\phi_k}^B$ appearing in the formula (\ref{E}) for 
the correlation coefficient $E(\theta_j,\phi_k)$ can be obtained from 
the Wigner function (\ref{Wignercond}) as a marginal distribution.
We have
\begin{equation}
P(x_{\theta_{j}}^A,x_{\phi_{k}}^B)=\int\limits_{-\infty}^{\infty}\int\limits_{-\infty}^\infty 
W_{c,AB}(S_{\mathrm{sh}}^Tr_{\theta_j,\phi_k})dp_{\theta_j}^A dp_{\phi_k}^B,
\label{Pmarginal}
\end{equation}
where $r_{\theta_j,\phi_k}=[x_{\theta_j}^A,p_{\theta_j}^A,x_{\phi_k}^B,p_{\phi_k}^B]$
and the symplectic matrix 
$S_{\mathrm{sh}}=S_{\mathrm{PS},A}(\theta_j) \oplus S_{\mathrm{PS},B}(\phi_k)$
describes local phase shifts applied to modes A
and B that map the measured quadratures $x_{\theta_j}^A$  and $x_{\phi_k}^B$ 
onto the quadratures $x^A$ and $x^B$, respectively.

In order to express the result of the integration in Eq. (\ref{Pmarginal}) 
in a compact matrix notation, we re-order the elements of the vector $r_{\theta_j,\phi_k}$ 
as follows,
\begin{equation}
\left[
\begin{array}{c}
x_{\theta_j} \\
x_{\phi_k}   \\
p_{\theta_j} \\
p_{\phi_k}   \\
\end{array}
\right]
=\left[
\begin{array}{cccc}
1 & 0 & 0 & 0 \\
0 & 0 & 1 & 0 \\
0 & 1 & 0 & 0 \\
0 & 0 & 0 & 1 \\
\end{array}
\right]
\left[
\begin{array}{c}
x_{\theta_j} \\
p_{\theta_j} \\
x_{\phi_k}   \\
p_{\phi_k}   \\
\end{array}
\right]
\end{equation}
which defines a matrix $S_{\mathrm{hom}}$. 
After these algebraic manipulations, the four matrices $\Gamma_{j,AB}$ appearing in the
exponents in Eq. (\ref{Wignercond}) transform to
\begin{equation}
\Gamma_{j,AB}'=S_{\mathrm{hom}}S_{\mathrm{sh}}\Gamma_{j,AB}S_{\mathrm{sh}}^TS_{\mathrm{hom}}^T
\equiv \left[
\begin{array}{cc}
A_j & C_j \\
C_j^T & B_j
\end{array}
\right],
\end{equation}
where we have divided the matrix $\Gamma_{j,AB}'$ into four sub-matrices with respect to
the $x$ vs $p$ splitting. A straightforward integration over $p_{\theta_j}^A$ 
and $p_{\phi_k}^B$  in Eq. (\ref{Pmarginal}) then yields the joint probability 
distribution,
\begin{equation}
P(x_{\theta_{j}}^A,x_{\phi_{k}}^B)=
\frac{1}{\pi P_{G} \sqrt{\det\gamma_{\mathrm{out}}} }
\sum_{j=1}^4 \frac{q_j e^{-y^T\Gamma_j y}}{\sqrt{\det \Gamma_{j,CD}}\sqrt{\det B_j}},
\label{Pjoint}
\end{equation}
where $y=(x_{\theta_j}^A,x_{\phi_k}^B)^T$ and
\begin{equation}
\Gamma_j=A_j-C_jB_j^{-1}C_j^T.
\end{equation}
Taking into account the choice of  binning, 
the normalization of the joint probability distribution, and its symmetry, 
$P(x_{\theta_{j}}^A,x_{\phi_{k}}^B)=P(-x_{\theta_{j}}^A,-x_{\phi_{k}}^B)$,
we can express the correlation coefficient as follows,
\begin{equation}
E(\theta_j,\phi_k)=4\int_0^{\infty}\int_0^{\infty}
P(x_{\theta_{j}}^A,x_{\phi_{k}}^B)dx_{\theta_{j}}^A dx_{\phi_{k}}^B-1.
\end{equation}
This last integral can be easily evaluated analytically. For a given $\Gamma_j$ matrix 
\begin{equation}
\Gamma_j=
\left[
\begin{array}{cc}
a_j & c_j \\
c_j & b_j
\end{array}
\right],
\end{equation}
the integral of the exponential term 
\begin{equation}
G_j=\int_0^{\infty}\int_0^{\infty} e^{-a_j y_1^2-b_jy_2^2-2c_jy_1 y_2 }dy_1 dy_2
\label{Gj}
\end{equation}
can be calculated by transforming to polar coordinates and integrating first 
over the radial coordinate and then over the angle.
After some algebra, we finally  arrive at
\begin{equation}
G_j=\frac{1}{2\sqrt{a_jb_j-c_j^2}}\left[
\frac{\pi}{2}-\arctan\frac{c_j}{\sqrt{a_jb_j-c_j^2}}\right].
\end{equation}
The final fully analytical formula for the correlation coefficient reads
\begin{equation}
E(\theta_j,\phi_k)=
\frac{4}{\pi P_{G} \sqrt{\det\gamma_{\mathrm{out}}} }
\left[\sum_{j=1}^4 \frac{q_jG_j}{\sqrt{\det \Gamma_{j,CD}}\sqrt{\det B_j}}\right]-1
\end{equation}
and the Bell factor can be expressed as 
\begin{equation}
S=E(\theta_1,\phi_1)+E(\theta_1,\phi_2)+E(\theta_2,\phi_1)-E(\theta_2,\phi_2).
\end{equation}

\subsubsection{Violation of Bell-CHSH Inequalities}

\begin{figure}[b]
\centerline{\resizebox{0.85\hsize}{!}{\includegraphics*{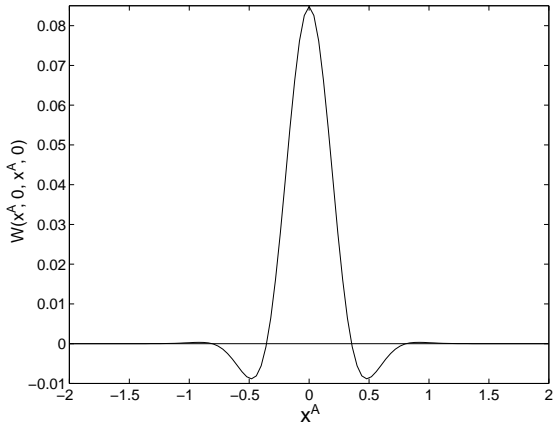}}}
\caption{A one-dimensional cut of the Wigner function of the two-mode state $\rho_{c,AB}$
along the line $x^B=x^A$, $p^A=p^B=0$ for $\lambda =0.6$ and beam splitters 
BS$_A$ and BS$_B$ transmittances
$T=0.95$. Notice the regions where $W$ is
negative.}
\label{Bell-Wigner}
\end{figure}
\par

A necessary condition  for the observation of a violation of Bell inequalities 
with homodyne detectors is that the Wigner function of the two-mode state 
used in the  Bell test is not positive definite. 
Figure \ref{Bell-Wigner} illustrates that the Wigner function (\ref{Wignercond}) 
of the conditionally generated state $\rho_{c,AB}$  is indeed negative 
in some regions of the phase space. The area of negativity, as well as the attained
negative values of W, are rather small, which indicates that we should not 
expect a high Bell violation with homodyning. 

\begin{figure}[t]
\centerline{\resizebox{0.99\hsize}{!}{\includegraphics*{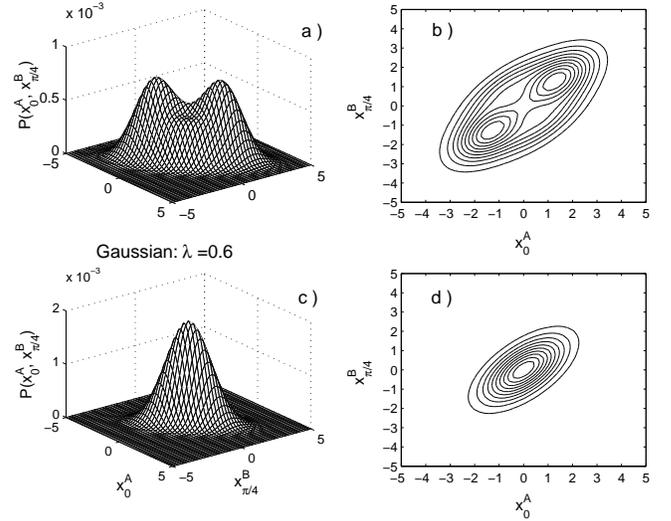}}}
\caption{Joint probability distribution $P(x_{\theta_{j}}^A,x_{\phi_{k}}^B)$.
Panels (a) and (b) show the distribution for the conditionally-prepared 
non-Gaussian state with $T=0.99$. Panels (c) and (d) display the 
distribution for the initial Gaussian two-mode squeezed vacuum state.
The curves are plotted for perfect detectors $\eta_{\mathrm{PD}}=\eta_{\mathrm{BHD}}=100$\%,
squeezing $\lambda=0.6$ and $\theta_{\mathrm{Alice}}=0$ and $\phi_{\mathrm{Bob}}=\pi/4$.}
\label{Bell-Joint}
\end{figure}

As we have shown in  Section II, the maximum Bell factor $S$ achievable 
with our setup and  sign binning is about $S=2.048$. We conjecture that 
this binning is optimal or close to optimal.  This is supported by the simple 
structure of the  joint probability distribution (\ref{Pjoint}). 
As can be seen in  Fig.~\ref{Bell-Joint}(a,b), $P$ exhibits two peaks, 
both located in the quadrants where  Alice's and Bob's measured quadratures 
have the same sign. Note also that the
two-peak structure is a clear signature of the non-Gaussian character of the state 
(c.f. Fig.~\ref{Bell-Joint}(c,d)).
We have carried out numerical calculations of $S$ for several other possible binnings 
which divide the quadrature axis into three or four intervals, and have not 
found any binning which would provide higher $S$ than the sign binning.
We have also performed optimization over the angles $\theta_j$ and $\phi_k$ and all 
the results and figures presented in this Section  were obtained for the optimal 
choice of angles $\theta_1=0$, $\theta_2=\pi/2$, $\phi_1=-\pi/4$, $\phi_{2}=\pi/4$.

\begin{figure}[t]
\centerline{\resizebox{0.95\hsize}{!}{\includegraphics*{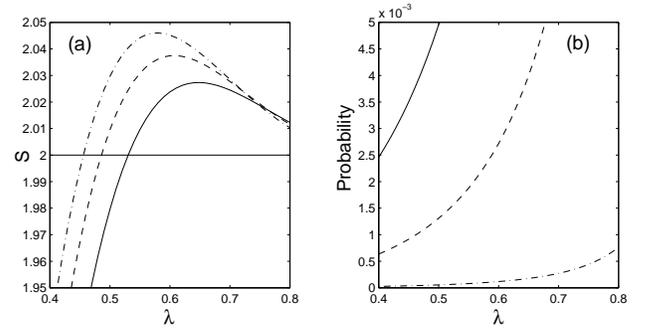}}}
\caption{Violation of Bell-CHSH inequality with the conditionally-prepared non-Gaussian state.
(a) Bell factor $S$ as a function of the squeezing.
(b) Probability of success of the generation of the non-Gaussian state as a function 
of the squeezing. The curves are plotted for perfect detectors 
($\eta_{\mathrm{\mathrm{PD}}}=\eta_{\mathrm{\mathrm{BHD}}}=100$\%) 
with $T=0.9$ (solid line), $T=0.95$ (dashed line), and $T=0.99$ (dot-dashed line). }
\label{Bell-S}
\end{figure}

Figure \ref{Bell-S}(a) illustrates that the Bell-CHSH inequality $|S|\le 2$ 
can  be violated with the proposed set-up, and shows that there is an 
optimal squeezing $\lambda_{\mathrm{\mathrm{opt}}}$ which maximizes $S$. 
This optimal squeezing is well predicted by the simple model
assuming perfect detectors with single-photon resolution (section II B), 
$\lambda_{\mathrm{\mathrm{opt}}}T\approx 0.57$. The curve plotted for $T=0.99$ 
practically coincides with the results obtained from the simple model presented 
in Sec. II B, c.f. Fig. 2(a). This confirms that in the limit $T\rightarrow 1$ the
detectors with single-photon sensitivity become for our purposes equivalent to
photodetectors with single-photon resolution. The maximum Bell factor achievable 
with our scheme is about $S_{\mathrm{max}}\approx 2.045$ which represents a violation 
of the Bell inequality by $2.2$\%. To get close to the $S_{\mathrm{max}}$ 
one needs sufficiently high (but not too strong)
squeezing. In particular, the value $\lambda\approx 0.57$ corresponds 
to approximately 5.6~dB of squeezing.
Figure \ref{Bell-S}(b) illustrates that
there is a clear trade-off between $S$ and the probability of success $P_{G}$. 
To maximize $S$ one should use highly transmitting beam splitters 
but this would reduce $P_G$. The optimal $T$ that should be chosen would 
clearly depend on the details of the experimental implementation.

\begin{figure}[t]
\centerline{\resizebox{0.95\hsize}{!}{\includegraphics*{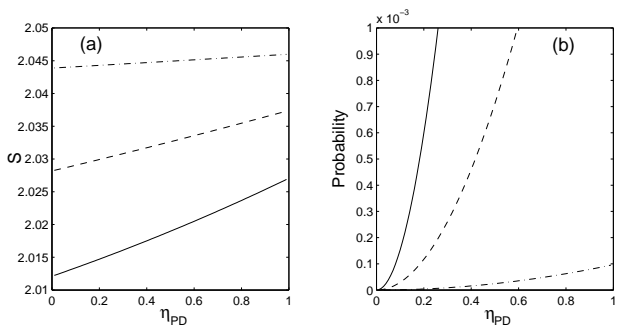}}}
\caption{Effect of the inefficiency of the photodetectors PD$_{A}$ and PD$_{B}$. 
(a) Bell parameter $S$ as a function of the efficiency $\eta_{\mathrm{PD}}$ of 
the photodetectors.
(b) Probability of success as a function of the efficiency $\eta_{\mathrm{PD}}$.
The curves are plotted for $T\lambda=0.57$, $\eta_{\mathrm{BHD}}=100\%$ and 
the same transmittances as in Fig. \ref{Bell-S}. }
\label{Bell-PD}
\end{figure}
\par

It follows from Fig.~\ref{Bell-PD}(a) that the Bell factor $S$ depends only very
weakly on the efficiency $\eta_{\mathrm{PD}}$ of the single-photon detectors, so the Bell
inequality can be violated even if $\eta_{\mathrm{PD}}\approx 1$\%. 
This is very important from the experimental point of view because, 
although the quantum detection efficiencies of the avalanche photodiodes may 
be of the order of $50$\%, the necessary spectral and
spatial filtering which selects the mode that is detected by the photodetector 
may reduce the overall detection efficiency to a few percent. 
Low detection efficiency only
decreases the probability of conditional generation $P_G$ of the non-Gaussian state,
see Fig.~\ref{Bell-PD}(b).  The dependence of $P_G$ on $\eta_{\mathrm{PD}}$ and $T$ can be very well 
approximated by a quadratic function, $P_{G}\approx \eta_{\mathrm{PD}}^2(1-T)^2$ which 
quickly drops when $\eta_{\mathrm{PD}}$ decreases. 
In practice, the  minimum necessary $\eta_{\mathrm{PD}}$ will be determined mainly by 
the constraints on the total time of the experiment and by the dark counts of the
detectors.

\begin{figure}[!t!]
\centerline{\resizebox{0.95\hsize}{!}{\includegraphics*{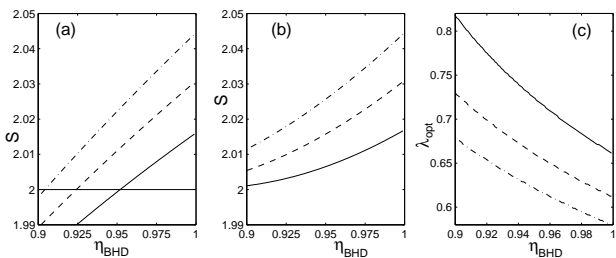}}}
\caption{Effect of inefficient homodyning.
(a) Bell parameter $S$ as a function of the efficiency $\eta_{\mathrm{BHD}}$ of 
the homodyning. The curve is plotted for $T\lambda=0.57$, $\eta_{\mathrm{PD}}=30\%$ and 
the same transmittances as in Fig. \ref{Bell-S}. 
(b) Bell parameter achieved for the optimal squeezing  $\lambda_{\mathrm{opt}}$ is plotted 
as a function of $\eta_{\mathrm{BHD}}$.
(c) Optimal squeezing $\lambda_{\mathrm{opt}}$ is plotted as a function of $\eta_{\mathrm{BHD}}$.
The curve is plotted for $\eta_{\mathrm{PD}}=30\%$ and 
the same transmittances as in Fig. \ref{Bell-S}. }
\label{Bell-BHD}
\end{figure}
\par

In contrast, the Bell factor $S$ strongly depends on the efficiency  
of the homodyne detectors, and $\eta_{\mathrm{\mathrm{BHD}}}$ must be above $\sim 90$\% 
in order to observe Bell violation, see Fig. \ref{Bell-BHD}. 
However, this is not an obstacle because 
such (and even higher) efficiency has been already achieved experimentally 
(see {\it e.g.} \cite{zhang03}). 
Interestingly, we have found that it is possible  to partially compensate for
imperfect homodyning with efficiency $\eta_{\mathrm{BHD}}<1$ by increasing the squeezing of 
the initial state.  This effect is illustrated in Fig. \ref{Bell-BHD}(b) which
shows the dependence of the Bell factor $S$ on $\eta_{\mathrm{BHD}}$ for optimal 
squeezing $\lambda_{\mathrm{opt}}$. Figure \ref{Bell-BHD}(c) then shows how the optimal 
squeezing increases with decreasing $\eta_{\mathrm{BHD}}$.

\begin{figure}[!t!]
\centerline{\resizebox{0.95\hsize}{!}{\includegraphics*{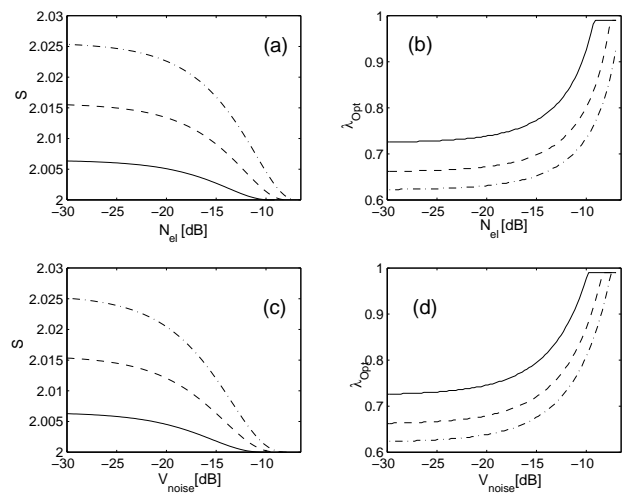}}}
\caption{Effect of the electronic noise and thermal input states.
(a) Maximum achievable Bell parameter $S$ with the optimal squeezing $\lambda_{\mathrm{opt}}$
as a function of the electronic noise $N_{\mathrm{el}}$.
(b) Optimal squeezing $\lambda_{\mathrm{opt}}$ giving the highest
Bell parameter $S$ for a given electronic noise. 
(c) Maximum Bell parameter $S$ as a function of the thermal noise 
of the input state $V_{\mathrm{noise}}$.
(d) Optimal squeezing $\lambda_{\mathrm{opt}}$ giving the highest
Bell parameter $S$ for a given thermal noise at the input.
The curves are plotted for  $\eta_{\mathrm{\mathrm{PD}}}=30$\%, 
$\eta_{\mathrm{\mathrm{BHD}}}=95$\%, and  $T=0.9$ (solid line), 
$T=0.95$ (dashed line), and $T=0.99$ (dot-dashed line).
}
\label{Bell-Noise}
\end{figure}
In addition to imperfect detection efficency $\eta_{\mathrm{BHD}}$, the electronic noise of the
homodyne detector is another factor that may reduce the observed Bell violation. 
We model the added electronic noise by assuming that
the effective quadrature that is detected $x_{\mathrm{det}}$ is related 
to the signal quadrature $x_S$ by a formula,
\[
x_{\mathrm{det}}=\sqrt{\eta_{\mathrm{\mathrm{BHD}}}} \, x_S
+\sqrt{1-\eta_{\mathrm{\mathrm{BHD}}}} \, x_{\mathrm{vac}}
+\sqrt{N_{\mathrm{el}}} \, x_{\mathrm{noise}},
\]
where $x_{\mathrm{vac}}$ and $x_{\mathrm{noise}}$ are two independent 
Gaussian distributed quadratures with zero mean and variance $1/2$, 
and $N_{\mathrm{el}}$ is the electronic noise variance expressed in 
shot noise units. On the level of covariance matrices, 
$N_{\mathrm{el}}$ can be included by modifying the formula for the noise matrix $G$ ,
\begin{equation}
G=(1-\eta_{\mathrm{\mathrm{BHD}}}+N_{\mathrm{el}})I_{AB}\oplus (1-\eta_{\mathrm{PD}})I_{CD} .
\end{equation}
The homodyne detector with electronic noise is actually equivalent to a detector without
noise but with a lower homodyne detector efficiency 
$\eta_{\mathrm{\mathrm{BHD}}}'=\eta_{\mathrm{\mathrm{BHD}}}/(1+N_{\mathrm{el}})$.
This can be shown by noting that the re-normalized quadrature 
$x_{\mathrm{det}}/\sqrt{1+N_{\mathrm{el}}}$  
is exactly a quadrature that would be detected by a balanced homodyne detector 
with $N_{\mathrm{el}}=0$ and efficiency $\eta_{\mathrm{\mathrm{BHD}}}'$. 
Our calculations reveal that the electronic noise 
should be $15-20$ dB below shot noise (see Fig. \ref{Bell-Noise}(a) and (b)), 
which is currently attainable with low-noise charge amplifiers. Again, higher squeezing can
partially compensate for the increasing noise. 
\par

So far we have assumed that the source in Fig. 1 emits pure two-mode squeezed vacuum
state. However, experimentally, it is very difficult to generate pure squeezed vacuum 
saturating the Heisenberg inequality. It is more realistic to consider a mixed Gaussian
state such as squeezed thermal state which can be equivalently represented by adding 
quadrature independent Gaussian noise with variance $V_{\mathrm{noise}}$ to each mode 
of the two-mode squeezed vacuum. The effect of the added noise stemming from  
input mixed Gaussian state is quite similar to the influence of the electronic noise 
of the homodyne detector,  see Fig. \ref{Bell-Noise} (c) and (d).
We find again that the added noise in the initial Gaussian 
state should be $15-20$~dB below  the shot noise.

\subsection{Four photon subtractions}

Until now we have focused on a single-photon subtraction
on each side (one photon removed from mode $A$ and one from mode $B$). 
If we now consider a scheme where two photons are subtracted from each mode, 
the de-gaussification of the state will be stronger 
and we may expect a higher Bell violation than before. To subtract two photons
from each mode, we only need to add one more unbalanced beam splitter
and photodetector on each side in Fig. 1. A successful state generation would
be indicated by simultaneous clicks of all four detectors. Assuming perfect
photon-number resolving detectors, the state generated from two-mode squeezed
vacuum (\ref{psiin}) by subtracting two photons from each mode can be expressed as
\begin{eqnarray}
|\psi_{\mathrm{out}}\rangle_{AB}&\propto& a^{2} b^{2}|\psi_{\mathrm{in}}(T^2\lambda)\rangle_{AB}
\nonumber \\
&\propto&\sum_{n}(n+2)(n+1)(T^2\lambda)^n |n,n\rangle_{AB},
\label{psioutfour}
\end{eqnarray}
and the probability of success reads
\begin{equation}
P_{\mathrm{4ph}}=2T^2(1-T)^4 \lambda^4 (1-\lambda^2)
\frac{1+10T^4\lambda^2+T^8\lambda^4}{(1-T^4\lambda^2)^5}.
\end{equation}
Since the state (\ref{psioutfour}) exhibits perfect photon number correlations, the Munro's
formula for the Bell factor can again be directly applied \cite{Munro99}. 
Numerical calculations show that the maximum Bell violation with 
the state (\ref{psioutfour}) and sign binning of quadratures 
is achieved for $T^2\lambda=0.40$ which yields $S_{\mathrm{max,4ph}}=2.064$,
which is indeed higher than the maximum achievable with two-photon subtraction,
$S_{\mathrm{max,2ph}}=2.048$, and very close to the maximum value $S=2.076$ \cite{Munro99}.

\begin{figure}[t]
\centerline{\resizebox{0.95\hsize}{!}{\includegraphics*{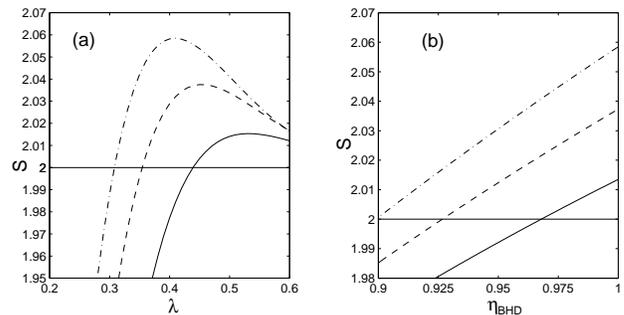}}}
\caption{
Violation of Bell-CHSH inequality with four photon subtractions.
(a) Bell parameter $S$ as a function of the squeezing $\lambda$ for perfect detectors 
$\eta_{\mathrm{PD}}=\eta_{\mathrm{BHD}}=100$\%.
(b) Bell parameter $S$ as a function of the efficiency $\eta_{\mathrm{BHD}}$ of 
the homodyning. The curve is plotted for $T^2\lambda=0.40$, $\eta_{\mathrm{PD}}=100\%$ and 
the same transmittances as in Fig. \ref{Bell-S}.  
}
\label{Bell-4CM}
\end{figure}

A more realistic description of the four-photon subtraction scheme that takes
into account realistic imperfect detectors can be developed using the approach
described in detail in Sec. IIIB. We find that the Wigner function of the 
conditionally generated state is a linear combination of sixteen Gaussians.
The results of numerical calculations are 
shown in Figs. \ref{Bell-4CM}(a) and (b), which illustrate that
the two-photon subtraction  from each mode yields higher violation of 
Bell-CHSH inequality than  one-photon subtraction only 
for very high transmittances $T>0.95$. For lower transmittances, the fact
that the photodetectors do not distinguish the number of  photons
reduces the Bell factor. 
Moreover, adding a second stage of photon subtractions dramatically decreases 
the probability of  generating the non-Gaussian state.
The probability can be estimated as $P_{G}\approx \eta_{\mathrm{PD}}^4(1-T)^4$, so for 
$T> 0.95$ and $\eta=50$\% we get $P_G \approx 10^{-6}$ and 
the duration of data acquisition would make the experiment infeasible.
We conclude that from the practical point of view there seems to be no 
advantage in using the scheme with four photon subtractions instead of the 
much simpler scheme with two photon subtractions.

\begin{figure}[t]
\centerline{\resizebox{0.7\hsize}{!}{\includegraphics*{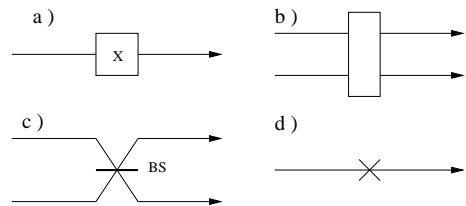}}}
\caption{Symbol convention.
(a) Single-mode squeezer along the $x$ quadrature.
(b) Two-mode squeezer.
(c) Beam splitter.
(d) Conditional subtraction of a photon as described in the preceding section.
}
\label{symbol}
\end{figure}

\section{Alternative schemes}

In this section we will study the violation of Bell-CHSH inequalities 
for a large group of alternative schemes, which involve  from one to
four photon subtractions. The main objective of this section is to compare 
the maximum Bell-CHSH factor $S$ obtained for the different proposed setups.
As the main purpose of this section is the comparison of the different schemes,  
we will consider only idealized schemes with almost perfect single-photon
subtraction on the beam splitters ($T=0.99$), 
and perfect photodetectors and homodyning ($\eta_{\mathrm{PD}}=\eta_{\mathrm{BHD}}=100$\%).
The maximum achievable Bell factor for each scheme presented below 
was determined by optimizing over the  angles $\theta_{1,2}$, $\phi_{1,2}$
as well as over the squeezing $\lambda$ of the initial Gaussian states. 
The sign binning of the measured quadratures has been used in all cases. 
All the schemes presented in this section  use the symbol 
convention depicted in Fig. \ref{symbol}.

\begin{figure}[t]
\centerline{\resizebox{0.7\hsize}{!}{\includegraphics*{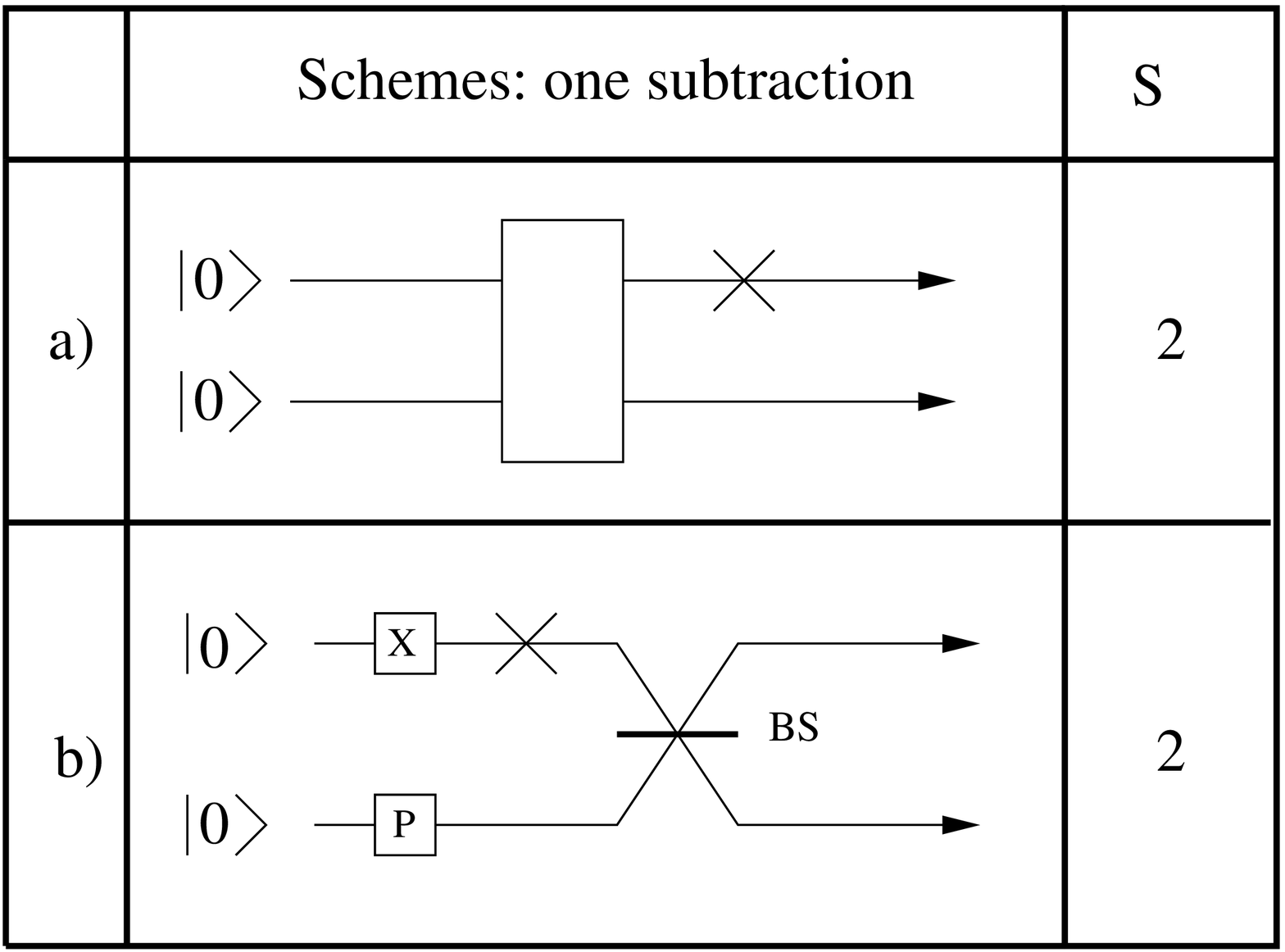}}}
\caption{Schemes with only one photon subtraction. The first column labels
the different setups proposed, the second shows the scheme and finally 
the last column gives the maximal Bell factor $S$.
(a) Photon subtraction after the creation of the two-mode squeezed vacuum.
(b) Photon subtraction  before mixing  two single-mode squeezed states on a
 beam splitter.
}
\label{Scheme1CM}
\end{figure}

In the preceding section, we have seen that
the probability of successful generation of a non-Gaussian 
state decreases significantly with the number of photon subtractions.
At the same time the complexity of the implementation of the experimental setup
increases with the number of photon subtractions. 
It is then obvious that the most interesting schemes for a Bell-CHSH violation 
are those  involving only one photon subtraction. Unfortunately, for the schemes
that we have considered (see Fig. \ref{Scheme1CM}), no violation is observed.

\begin{figure}[t]
\centerline{\resizebox{0.7\hsize}{!}{\includegraphics*{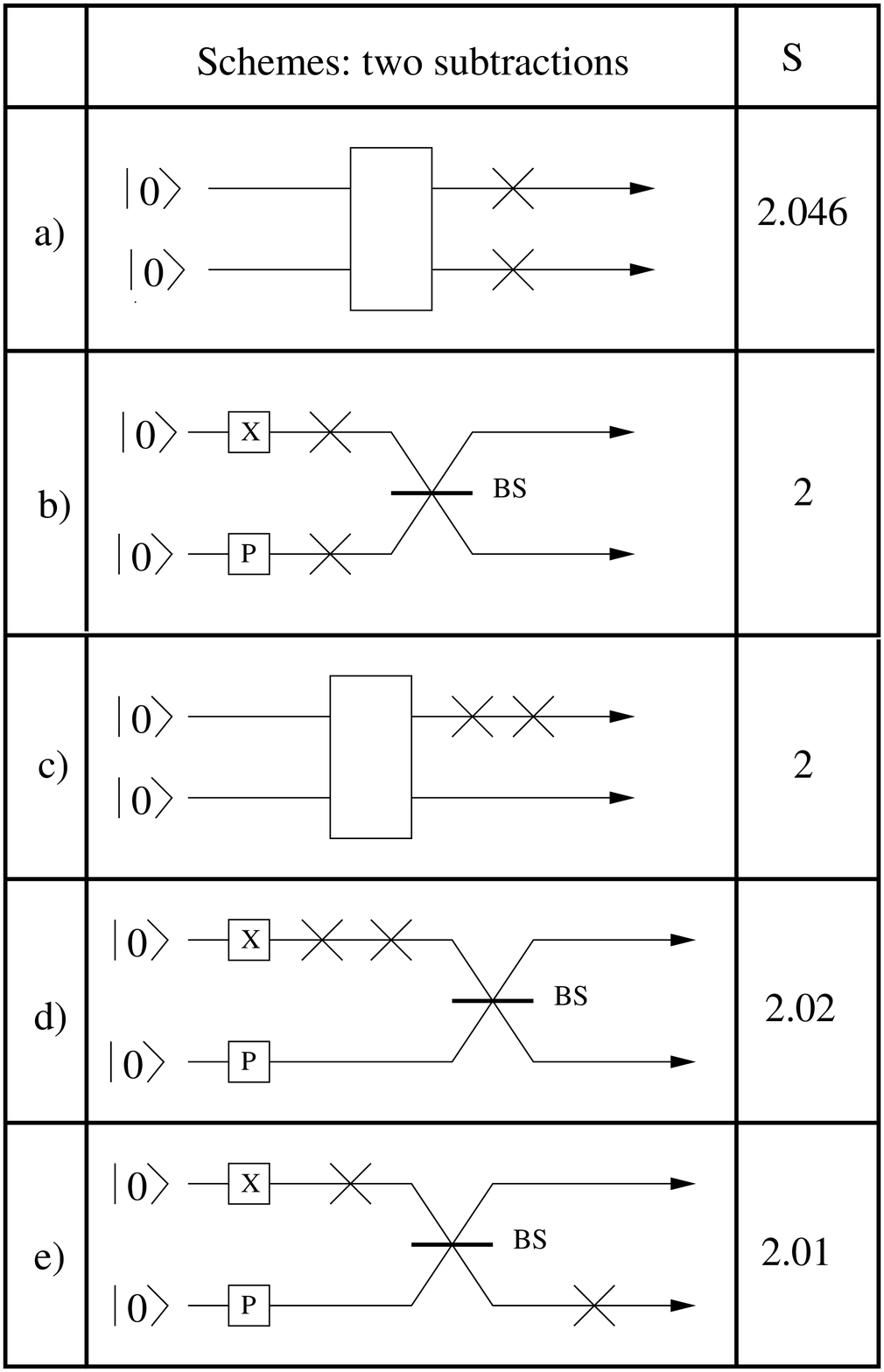}}}
\caption{
Schemes with two photon subtractions. 
The right column gives the maximal value of the Bell
factor $S$ for the proposed setups.}
\label{scheme2CM}
\end{figure}

After one photon subtraction, the simplest schemes are those with two photon 
subtractions. In the preceding sections it was shown that it is possible to
violate Bell-CHSH inequality with
two photon subtractions (scheme Fig. \ref{scheme2CM}(a)).
It follows from  Fig. \ref{scheme2CM}
that several other schemes (scheme Fig. \ref{scheme2CM}(d) and (e)) 
also violate Bell-CHSH inequality but the maximal achievable 
Bell factor $S$ is much smaller in comparison to the scheme shown in 
Fig. \ref{scheme2CM}(a).     

By adding one more photon subtraction to the schemes 
shown in Fig. \ref{scheme2CM},
we can construct an ensemble of schemes with three photon subtractions. 
After numerical optimization  we have found that none of these schemes 
succeeds to violate Bell-CHSH inequality. This striking result together
with the the fact that we have not found any violation for schemes based on a
single subtraction suggests that it may be necessary to have a scheme with an
even number of photon subtractions in order to observe $S>2$.

\begin{figure}[t]
\centerline{\resizebox{0.7\hsize}{!}{\includegraphics*{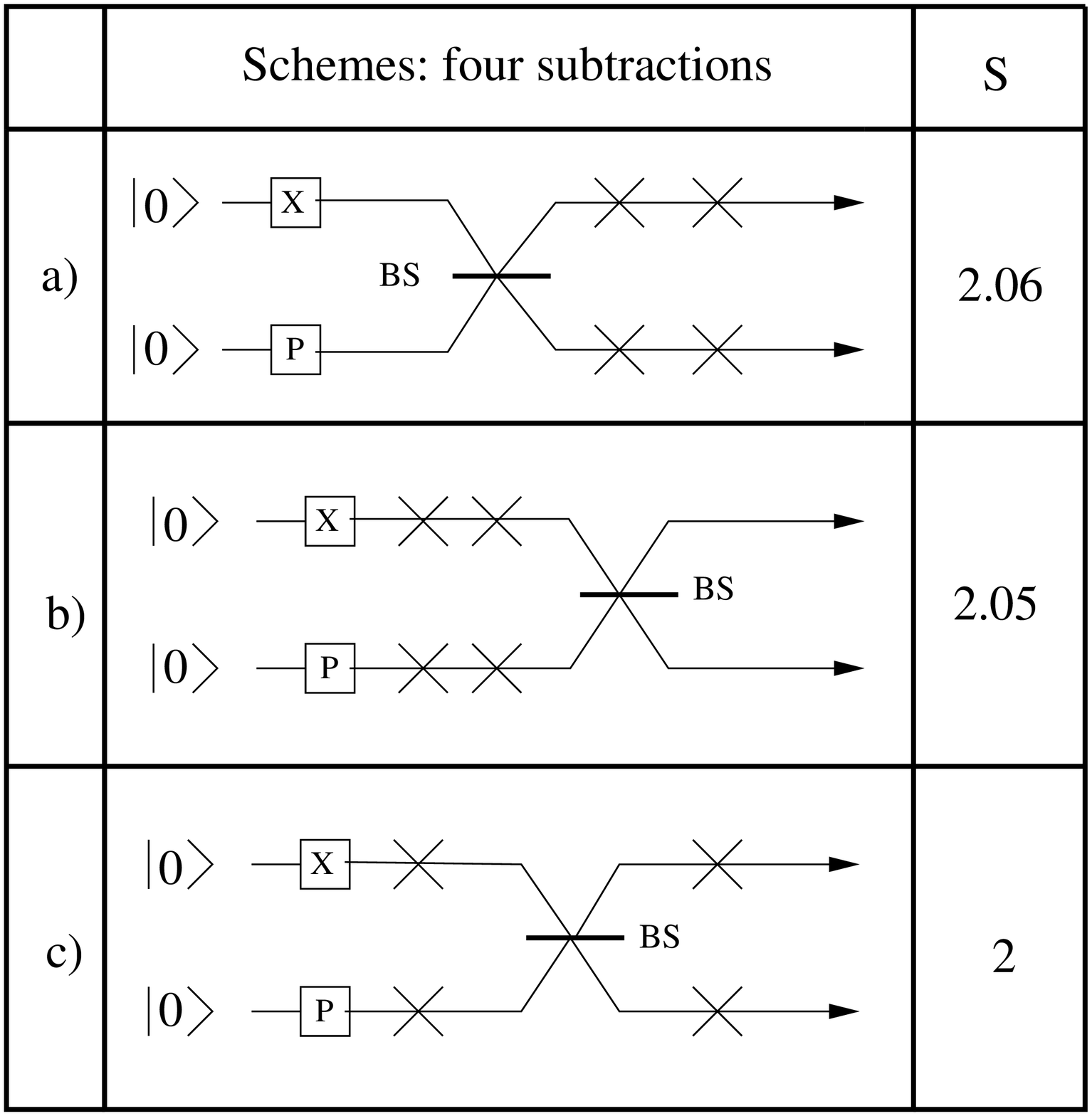}}}
\caption{
Schemes with four photon subtractions.  Last column gives the maximal value of the Bell
factor $S$ for the proposed setups.
}
\label{scheme4CM}
\end{figure}

In the preceding section, we have also proposed one scheme with four 
photon subtractions that violates Bell-CHSH inequality. Many other possible 
schemes exist where four photons are subtracted. Figure \ref{scheme4CM} 
illustrates some particular examples, which are based
on the preparation of two-mode squeezed vacuum
via mixing of two single-mode squeezed states on an balanced beam splitter.
The photon subtractions are symmetrically placed to both modes. Strikingly, if
all four photons are subtracted either before or after mixing on a beam
splitter, then we get $S>2$. However, if a single photon is subtracted 
from each mode both before and after combining the modes on a beam splitter, 
then we do not obtain any Bell violation. 

\begin{figure}[!t!]
\centerline{\resizebox{0.85\hsize}{!}{\includegraphics*{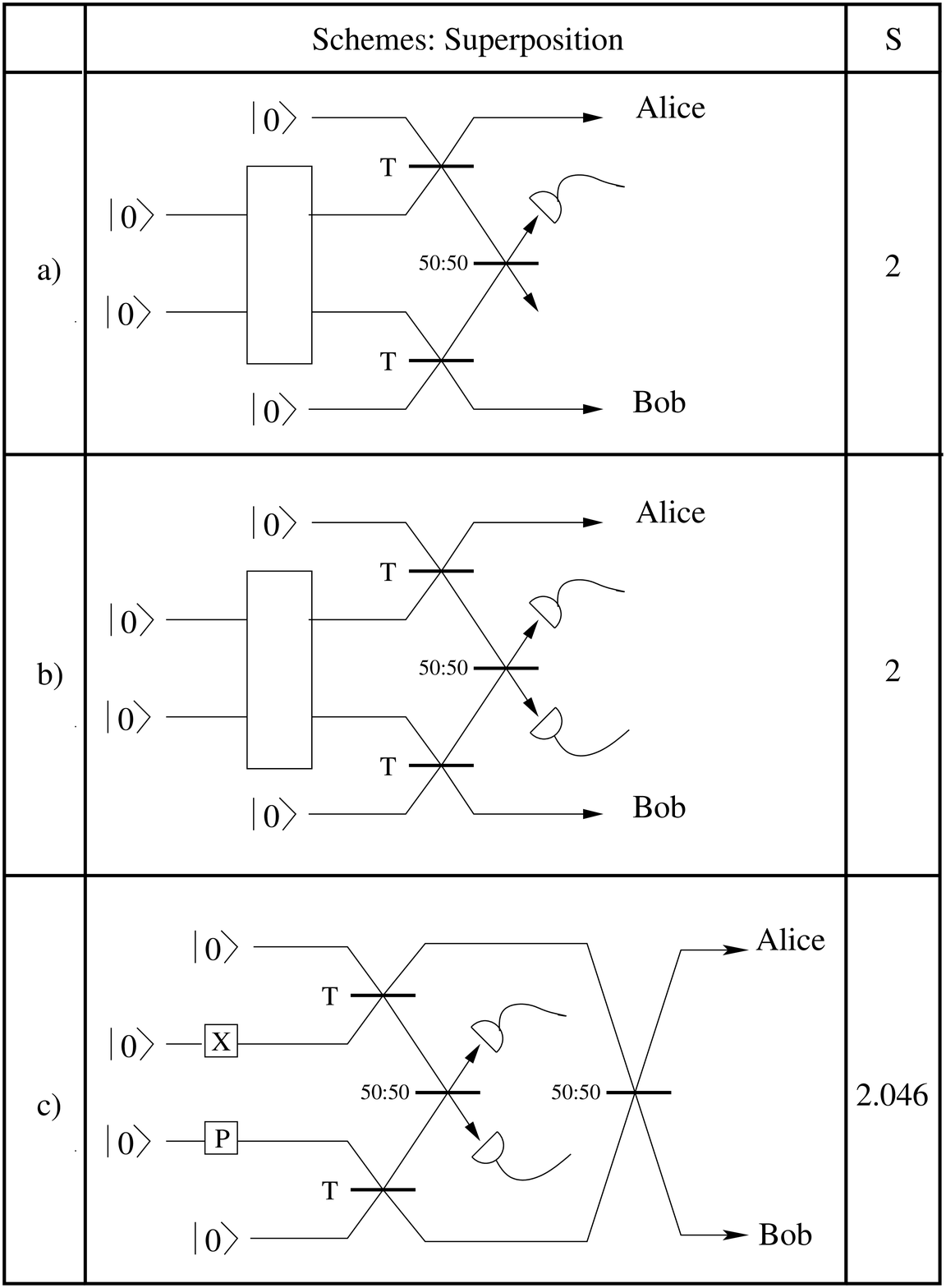}}}
\caption{Schemes consisting of superpositions of other schemes proposed above.
(a) Superposition of one photon subtraction on modes $A$ or $B$.
(b), (c) Superposition of two photon subtractions on modes $A$ or $B$.
}
\label{superpos}
\end{figure}

Finally we have also studied an alternative group of schemes where 
instead of subtracting photons separately from modes $A$ and $B$, 
we mix the auxiliary  modes $C$ and $D$  on a balanced beam splitter
before the detection on the photodetectors. Consider the scheme depicted 
in Fig. \ref{superpos}(a) where only a single photon is subtracted.
The mixing of modes $C$ and $D$ on a beam splitter erases the information 
about the origin of the detected photon which implies that the conditionally
prepared state is a coherent superposition of states where a single photon has
been removed either from mode $A$ or from mode $B$. However, even this 
modification does not lead to Bell violation with just a single subtraction. 

We can extend the scheme by placing a photodetector at both output
ports of the beam splitter, cf. Fig. \ref{superpos}(b).  
In the limit of a high transmittance $T\rightarrow 1$, the  conditioning 
on the click of each detector selects the events where there were altogether 
two photons at the beam-splitter inputs. 
The bosonic properties of the photons imply that a simultaneous click
of both photodetectors occurs only if the two subtracted photons 
are coming from the same mode ($A$ or $B$) \cite{Hong87}, but again we 
do not know from which mode the two photons are subtracted. 
This scheme is thus equivalent to the superposition of two schemes of the type
shown in   Fig. \ref{scheme2CM}(c). Unlike the scheme 
in  Fig. \ref{scheme2CM}(c), the scheme in Fig. \ref{superpos}(b) is
symmetric with respect to the modes $A$ and $B$. However, no violation can be
observed.  On the other hand, the scheme in Fig. \ref{superpos}(c) 
leads to $S>2$ by
realizing a superposition of states where two photons are subtracted from a
single-mode squeezed vacuum state and this state is then mixed with another
single-mode squeezed vacuum on a balanced beam splitter, see \ref{scheme2CM}(d). 
In comparison to the scheme in Fig. \ref{scheme2CM}(d), we obtain 
much higher violation $S=2.046$. 

\par

\section{Conclusions}

We have proposed an experimentally  feasible setup allowing 
for a loophole-free Bell test with efficient homodyne detection using 
a non-gaussian entangled state generated from a two-mode squeezed vacuum state
by subtracting a single photon from each mode. We have presented 
a full analytical description of  a realistic setup with imperfect detectors,
noise and mixed input states. We have studied in detail  the influence of
the detector inefficiencies, the electronic noise of homodyne detector,
and the input mixed states, on the achievable Bell violation. The main 
feature of the present scheme is that it is largely insensitive 
to the detection efficiency of the avalanche photodiodes that are used for
conditional preparation of the non-gaussian state, so that detector efficiencies of
the order of a few per cent are sufficient. On the other hand, the detection
efficiency of the balanced homodyne detector should be of the order of $90$\%
and the electronic noise of the homodyne detector should be at least $15$ dB
below the shot noise level. The optimal squeezing that yields maximum Bell
violation depends on the experimental circumstances but is, generally speaking, 
within the range of experimentally attainable values. As a rule, the optimal
squeezing increases with decreasing $\eta_{\mathrm{\mathrm{BHD}}}$ 
and increasing noise.

We have also discussed several alternative schemes that involve 
the subtraction of one, two, three or four photons. The experimentally simplest
and most appealing schemes are those  where only a single photon is subtracted
because photon subtraction is a delicate operation and also each subtraction
in the scheme drastically reduces the probability of successful state generation. 
Unfortunately, we have not been able to find a scheme with only a single
subtraction which would exhibit violation of Bell inequalities. However,
the class of schemes that we have studied is still somewhat restricted. 
One can thus hope that such a scheme may be designed  by considering more 
complicated setups involving unbalanced beam splitters and possibly 
a different binning procedure \cite{Eisert04private}. This issue certainly 
deserves further investigation.

Among all the schemes where two photons are subtracted, the maximum 
violation $S=2.046$ is achieved by the scheme discussed in Sections II and III.
Taking into account that we have not found any  scheme with three photon 
subtractions which would violate Bell-CHSH inequality, 
the only way of exceeding the 2.046 violation appears to be  by subtracting 
four photons. This scheme has been analyzed in some detail in Sec. IIIC 
where it was shown that this  allows us to reach the Bell factor $S=2.06$.
Unfortunately, the price to pay for this slight increase of $S$ is that 
the probability of successful conditional generation 
is so low that it makes the experiment infeasible.

The results presented in this paper provide a clear example of the utility 
of conditional photon subtraction which can be considered as an important
novel tool in quantum optics and quantum information processing with 
continuous variables. Besides violation of
Bell inequalities, this method can be used to generate highly non-classical
states of light \cite{Wenger04}, to improve the fidelity of teleportation 
of continuous variable states \cite{Opatrny00,Cochrane02,Olivares03} 
and it forms a key ingredient of recently proposed entanglement
purification protocols for continuous variables \cite{Browne03b,Eisert04}. 
The very recent experimental
demonstration of a single photon subtraction from a single-mode squeezed vacuum
state provides a strong incentive for further theoretical and experimental
developments along these lines, and we can thus expect that some of the 
schemes discussed in the present paper will be experimentally implemented 
in a not too distant future. 

\emph{Note added:} After this work was completed we have learned that a 
scheme for observing a violation of Bell inequalities similar to the scheme 
discussed in Sec. II of the present paper has been independently proposed 
by H. Nha and H.J. Carmichael \cite{Nha04}.

\acknowledgments

We would like to thank Ph. Grangier, R. Tualle-Brouri, J. Wenger, and J. Eisert
for many stimulating discussions.
We acknowledge financial support from the Communaut\'e Fran\c{c}aise de
Belgique under grant ARC 00/05-251, from the IUAP programme of the Belgian
government under grant V-18, from the EU under projects RESQ
(IST-2001-37559) and CHIC (IST-2001-33578). JF also acknowledges support
from the  grant LN00A015  of the Czech Ministry of Education.

\end{document}